\documentclass[seceq]{ptptex}

\usepackage{graphicx}

\title{
Two or Four: A Hint from Scalar Mesons in 
Radiative $\phi$ Decays~?~\footnote{%
Talk given by M.~Harada at 
Yukawa International Seminar (YKIS) 2006
``New Frontiers in QCD  --Exotic Hadrons and Hadronic Matter--''.
 } %
}

\author{
Deirdre \textsc{Black}$^{1,}$
Masayasu \textsc{Harada}$^{2,}$
and
Joseph \textsc{Schechter}$^{3,}$
}

\inst{
$^{1}$ University of Cambridge, Department of Physics,
Cavendish Laboratory,\\
J J Thomson Avenue
Cambridge CB3 0HE, UK\\
$^{2}$ Department of Physics, 
Nagoya University, Nagoya 464--8602, Japan\\
$^{3}$ Department of Physics, Syracuse University
Syracuse, NY 13244-1130, USA
}

\abst{
In this write-up,
we summarize our recent analysis of radiative decays involving
light scalar mesons.
Our analysis using the vector meson dominance model at tree level
indicates that it may be difficult to distinguish $qq\bar{q}\bar{q}$
picture and $q\bar{q}$ picture for the light scalar nonet.
Our result on the process of $\phi \rightarrow \pi^0 \eta \gamma$
shows that the derivative-type $f_0 K\bar{K}$ interaction
reproduces experimental data below 950\,GeV well, 
but gives a poor fit above 950\,GeV, i.e., in the energy region
around the mass of $a_0(980)$, but that
the discrepancy can be compensated by the effect of the $K$ loop.
}

\begin{document}

\maketitle

\section{Introduction}

According to recent theoretical and experimental analyses, there
is a strong possibility that nine light scalar mesons exist 
below 1 GeV, and
they form a scalar nonet~\cite{PDG}:
$\sigma(600)$ and $\kappa(900)$ together with the
well-established $f_0(980)$ and $a_0(980)$.
As is shown 
in Ref.~\citen{Black-Fariborz-Sannino-Schechter:99},
the mass hierarchy of the light scalar nonet can be
explained qualitatively when the members of the nonet have
a $qq\bar{q}\bar{q}$ quark structure.
The 4-quark scalar mesons have the same quantum numbers as
the ordinary scalar mesons made from the quark and 
anti-quark (2-quark picture).
The patterns of the interactions of the scalar mesons to other mesons
made from $q\bar{q}$, on the other hand, depend on the quark structure
of the scalar mesons.
The analysis of the interactions of the scalar mesons
will shed some light on the quark structure of the scalar nonet.
In 
Refs.~\citen{Black-Fariborz-Sannino-Schechter:99,Fariborz-Schechter},
several hadronic processes related to the scalar mesons are studied,
which shows that the scalar nonet takes dominantly the
$qq\bar{q}\bar{q}$ structure.

For getting more information on the structure
of the low-lying scalar mesons,
we studied the radiative decays involving scalar 
mesons in Refs.~\citen{BHS:1} and \citen{BHS:2}.
In this write-up we summarize the main results of the analyses.

\section{Scalar Nonet Field}

In Ref.~\citen{Black-Fariborz-Sannino-Schechter:99},
the scalar meson nonet is embedded into the $3\times3$ matrix
field $N$ as
\begin{equation}
N = 
\left(\begin{array}{ccc}
   \displaystyle \left(N_T+a_0^0\right) / \sqrt{2}
 & \displaystyle a_0^+
 & \displaystyle \kappa^+
\\
   \displaystyle a_0^-
 & \displaystyle \left(N_T-a_0^0\right) / \sqrt{2}
 & \displaystyle \kappa^0
\\
   \displaystyle \kappa^-
 & \displaystyle \bar{\kappa}^0
 & \displaystyle N_S
\\
\end{array}\right)
\ ,
\label{scalar nonet}
\end{equation}
where
$N_T$ and $N_S$ represent the ``ideally mixed'' fields.
The physical $\sigma(600)$ and $f_0(980)$ fields are expressed by
the linear combinations of these $N_T$ and $N_S$ as
\begin{equation}
\left(\begin{array}{c}
  \displaystyle \sigma \\ \displaystyle f_0 \\
\end{array}\right)
=
\left(\begin{array}{cc}
 \displaystyle \cos\theta_S & \displaystyle -\sin\theta_S \\ 
 \displaystyle \sin\theta_S & \displaystyle \cos\theta_S \\
\end{array}\right)
\,
\left(\begin{array}{c}
  \displaystyle N_S \\ \displaystyle N_T \\
\end{array}\right)
\ ,
\label{mixing}
\end{equation}
where $\theta_S$ is the scalar mixing angle.
The scalar mixing angle $\theta_S$
can parameterize the quark contents
of the scalar nonet field:
The case with $\theta_S = \pm 90^\circ$
is a natural assignment of the scalar meson nonet based on 
the $q\bar{q}$ picture, while
the case with $\theta_S = 0^\circ$ or $180^\circ$
is a natural assignment based on 
the $qq\bar{q}\bar{q}$ picture (see, e.g., 
Ref.~\citen{Black-Fariborz-Sannino-Schechter:99} for details).
Then, the present treatment of nonet field with the scalar mixing
angle can express both pictures for quark contents.

By fitting to the masses of the scalar nonet members,
the value of $\theta_S$ was 
found~\cite{Black-Fariborz-Sannino-Schechter:99}
to be
either $\theta_S \simeq - 20^\circ$ (corresponding to the case where
the scalar nonet is dominantly made from $qq\bar{q}\bar{q}$)
or $\theta_S \simeq - 90^\circ$ (corresponding to the case
where the scalar nonet is from $q\bar{q}$).
Some preference for
the $\theta_S= -20^\circ$ case for obtaining the best value of
the coupling constant $\gamma_{f\pi\pi}$, 
was expressed in section IV of 
Ref.~\citen{Black-Fariborz-Sannino-Schechter:99}.

\section{Radiative Decays Involving Light Scalar Mesons}

In Ref.~\citen{BHS:1}, 
the trilinear scalar-vector-vector terms were included
into the effective Lagrangian as
\begin{eqnarray}
{\cal L}_{SVV} &=&  \beta_A \,
\epsilon_{abc} \epsilon^{a'b'c'}
\left[ F_{\mu\nu}(\rho) \right]_{a'}^a
\left[ F_{\mu\nu}(\rho) \right]_{b'}^b N_{c'}^c
{}+
 \beta_B \, \mbox{tr} \left[ N \right]
\mbox{tr} \left[ F_{\mu\nu}(\rho) F_{\mu\nu}(\rho) \right]
\nonumber\\
&& 
{}+
 \beta_C \, \mbox{tr} \left[ N F_{\mu\nu}(\rho) \right]
\mbox{tr} \left[ F_{\mu\nu}(\rho) \right]
{}+
 \beta_D \, \mbox{tr} \left[ N \right]
\mbox{tr} \left[ F_{\mu\nu}(\rho) \right]
\mbox{tr} \left[ F_{\mu\nu}(\rho) \right]
\ ,
\label{SVV}
\end{eqnarray}
where $N$ is the scalar nonet field defined in 
Eq.~(\ref{scalar nonet}).
$F_{\mu\nu}(\rho)$ is the field strength of the
vector meson fields defined as
$ F_{\mu\nu}(\rho) = 
  \partial_\mu \rho_\nu - \partial_\nu \rho_\mu - i 
  \tilde{g} \left[ \rho_\mu \,,\, \rho_\nu \right]
$,
with $\tilde{g} \simeq4.04$~\cite{Harada-Schechter} being
the coupling constant.
In Ref.~\citen{BHS:1}, the vector meson dominance is assumed to be
satisfied in the radiative decays involving the scalar mesons.  Then,
the above Lagrangian (\ref{SVV}) determines all
the relevant interactions.
Actually, the $\beta_D$ term will not contribute so there
are only three relevant parameters $\beta_A$, $\beta_B$ and $\beta_C$.

We determined
the values of $\beta_A$ and $\beta_C$ 
from the experimental values of 
$\Gamma(a_0\rightarrow \gamma\gamma)$ and
$\Gamma(\phi \rightarrow a_0 \gamma)$, independently of the
scalar mixing angle $\theta_S$.
The value of $\beta_B$, on the other hand, depends on $\theta_S$
and there are two possible solutions for each $\theta_S$.
We determined the value for $\theta_S = -20^\circ$ as well as
for $\theta_S = -90^\circ$,
and then made several predictions on radiative decays
such as $\Gamma(\phi \rightarrow f_0\gamma)$ and
$\Gamma(\sigma \rightarrow \gamma\gamma)$.
We found~\cite{BHS:1}
that the predictions for two cases of the scalar
mixing angle are very close to each other (see Table~\ref{tab1}).
This result indicates that it may be difficult to
distinguish two pictures just from radiative decays.
Of course, other radiative decays 
should be studied to get more
information on the structure of the scalar mesons.
\begin{table}[htbp]
\begin{center}
\begin{tabular}{|l|rr|rr|}
\hline
$\theta_S$ & \multicolumn{2}{c|}{$-20^\circ$}
  & \multicolumn{2}{c|}{$-90^\circ$} \\
\hline
$\beta_A$ & $0.72\pm0.12$ & $0.72\pm0.12$ 
  & $0.72\pm0.12$ & $0.72\pm0.12$ \\
$\beta_B$ & $0.61\pm0.10$ & $-0.62\pm0.10$ 
  & $-0.12\pm0.13$ & $1.1\pm0.1$ \\
$\beta_C$ & $7.7\pm0.52$ & $7.7\pm0.52$ 
  & $7.7\pm0.52$ & $7.7\pm0.52$ \\
\hline
$\Gamma(\sigma\rightarrow\gamma\gamma)$ 
   & $0.024\pm0.023$ & $0.38\pm0.09$ 
   & $0.023\pm0.024$ & $0.37\pm0.10$ \\
$\Gamma(\phi\rightarrow\sigma\gamma)$ & $137\pm19$ & $33\pm9$ 
   & $140\pm22$ & $35\pm11$ \\
$\Gamma(\omega\rightarrow\sigma\gamma)$ & $16\pm3$ & $33\pm4$
   & $17\pm4$ & $33\pm5$ \\
$\Gamma(\rho\rightarrow\sigma\gamma)$ & $0.23\pm0.47$ & $17\pm4$
   & $0.20\pm0.43$ & $17\pm4$ \\
$\Gamma(f_0\rightarrow\omega\gamma)$ 
   & $126\pm20$ & $88\pm17$ 
   & $125\pm19$ & $86\pm16$ \\
$\Gamma(f_0\rightarrow\rho\gamma)$ & $19\pm5$ & $3.3\pm2.0$
   & $18\pm8$ & $3.4\pm3.2$ \\
$\Gamma(a_0\rightarrow\omega\gamma)$ & $641\pm87$ & $641\pm87$
   & $641\pm87$ & $641\pm87$ \\
$\Gamma(a_0\rightarrow\rho\gamma)$ & $3.0\pm1.0$ & $3.0\pm1.0$
   & $3.0\pm1.0$ & $3.0\pm1.0$ \\
\hline
\end{tabular}
\end{center}
\caption[]{Fitted values of $\beta_A$, $\beta_B$ and $\beta_C$
together with the predicted values of 
the decay widths of 
$V \rightarrow S + \gamma$ and $S \rightarrow V + \gamma$
for $\theta_S = -20^\circ$ and $\theta_S = -90^\circ$.
Units of $\beta_A$, $\beta_B$ and $\beta_C$ are $\mbox{GeV}^{-1}$
and those of the decay widths are keV.
}\label{tab1}
\end{table}

We should note that
our prediction on the $\phi \rightarrow f_0 \gamma$ is
too small when compared with experiment~\cite{PDG,recentexpts}.
A preliminary investigation including the effect of mixing
between a light non-$q\bar{q}$ scalar nonet $N$ and a heavier
$q\bar{q}$-type nonet $N^\prime$ described 
through a mixing Lagrangian 
${\cal L}_{\rm mix} = \gamma \mbox{Tr}\left( N N^\prime \right)$
introduced to explain the properties of the $I=1$ and $I=1/2$ scalar
meson in Ref.~\citen{BFS}
showed that the width is still too small.
Actually, the mixing between the scalar mesons, especially for
$I=0$ states are more complicated than just ${\cal L}_{\rm mix}$
(see, e.g., Ref.~\citen{mixing}).
An analysis in Ref.~\citen{TKM:05} shows that the inclusion of
the mixing of $f_0(980)$ to heavier $f_0$'s improves the prediction
but not enough for explaining experiment.

We should include the correction from the $K$-loop,
which gives an important contribution in the models
with non-derivative type $f_0 K \bar{K}$ 
interaction\cite{radiative}.
Our model includes the derivative-type $f_0 K \bar{K}$ interaction,
so that it is not so clear that the same mechanism appears in the
present model.
In Ref.~\citen{BHS:2} we studied the $K$-loop correction
to the effective $\phi$-$a_0$-$\gamma$ interaction, and studied
the process of $\phi \rightarrow \pi^0 \eta \gamma$.

\begin{figure}[htb]
\begin{minipage}{6cm}
\begin{center}
\includegraphics[width=5.0cm]{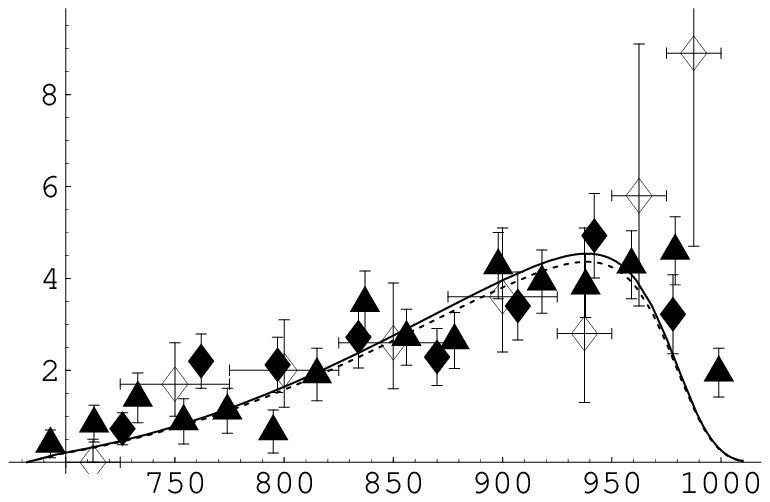}
\end{center}
\end{minipage}
\begin{minipage}{6cm}
\begin{center}
\includegraphics[width=5.0cm]{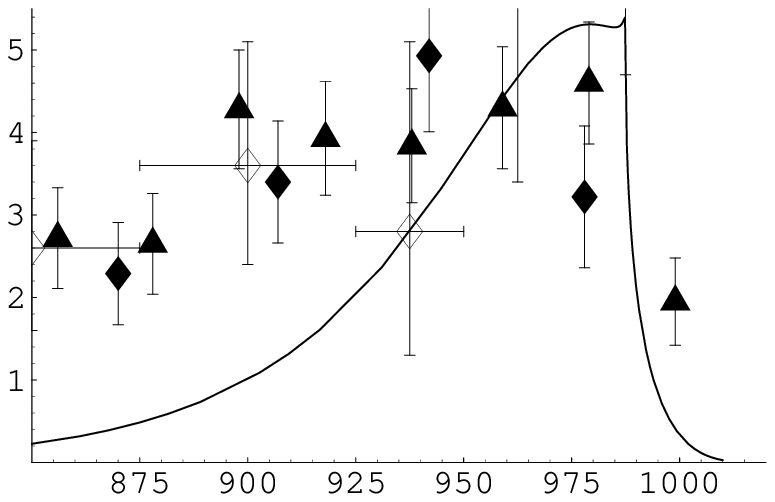}
\end{center}
\end{minipage}
\caption[]{
$d B(\phi\rightarrow \pi^0\eta \gamma) /d q \times
 10^7$ (in units of $\mbox{MeV}^{-1}$) as a function in
 the $\pi^0$-$\eta$ invariant mass $q = m_{\pi^0\eta}$ (in MeV).
Left panel shows the best fitted curve with
only the tree level contribution included,
while right panel shows the one with only 
$K$-loop contribution.
 Experimental data indicated by white diamonds ($\Diamond$)
 are from the SND collaboration~\cite{recentexpts}
 and those by filled triangles 
 and filled diamonds are shown in Ref.~\citen{Achasov-Kisilev}
 extracted from the KLOE collaboration~\cite{recentexpts}.
}\label{fig1}
\end{figure}
In the left panel of Fig.~\ref{fig1},
we show the best fitted curve for
$d B(\phi\rightarrow \pi^0\eta \gamma) /d q$ 
using the derivative-type $f_0 K\bar{K}$ interaction
together with experimental data~\cite{recentexpts,Achasov-Kisilev}.
This shows that our model with the derivative interaction 
well reproduces experimental data.
On the other hand, as seen in Fig.5 of Ref.~\citen{BHS:2},
if one were to use a tree model with nonderivative
SPP-type couplings, the resonant peaks were seen to get
completely washed out. This would appear to be an advantage
for the derivative coupling, which is dictated by chiral
symmetry in the present framework. 
Nevertheless, since
even with derivative coupling the spectrum shape is not
very well fitted in the energy region
above 950\,MeV,
there must be another mechanism at work.

In the right panel of Fig.~\ref{fig1}, we show the 
contribution from the $K$-loop.
This shows that,
for $q$ below the resonance region, the $K$-loop
contribution in the present model falls off rapidly, as one might
reasonably expect with derivative coupling, and lies lower
 than the data points.
In the energy region around the mass of $a_0(980)$, on the other hand,
the $K$-loop gives an important contribution.
A next step is to combine the tree and one-loop contributions
as well as the effect of mixing
between a $qq\bar{q}\bar{q}$-type scalar nonet and a 
$q\bar{q}$-type nonet~\cite{TMKwork}.
Determination of the mixing between two types of scalar nonets
is expected to give an important clue to understand 
the vacuum structure of underlying QCD~\cite{FJS}.

\vspace{-0.1cm}

\section*{Acknowledgements}

\vspace{-0.1cm}

The work of D.B. is supported in part by the Royal Society, UK. 
The work of M.H. is 
supported in part by the Daiko Foundation \#9099, 
the 21st Century
COE Program of Nagoya University provided by 
JSPS (15COEG01).
The work of J.S. is supported in part by the U. S. DOE
under contract No. DE-FG-02-85ER 40231.


\begin{thebibliography}{99}
\bibitem{TMKwork}
  Quite recently, an analysis with tree plus $K$-loop contributios
  are done in Ref.~\citen{TMK:07}, which shows that the data for
  $\phi \rightarrow \pi^0 \eta \gamma$ and 
  $\phi \rightarrow \pi \pi \gamma$ are well reproduced.
\end{thebibliography}
\end{document}